

\input nutex

\def\ub{\overline{u}}
\def\o{\omega}

\def\gb{\overline{g}}

\def\ub{\overline{u}}
\def\o{\omega}

\def\u#1{{\bf #1}}

\def\z{{\bf Z}^{p}}
\def\r{{\bf R}^{p}}
\def\c{\centerline}

\hoffset=0.3in
\voffset=0.4in

\baselineskip=24pt

\c{\bf The Chiral Potts Model and Its Associated Link Invariant}
\bigskip
\c{F. Y. Wu and P. Pant}
\c{Department of Physics}
\c{Northeastern University}
\c{Boston, Massachusetts, 02115}
\medskip
\c{C. King}
\c{Department of Mathematics}
\c{Northeastern University}
\c{Boston, Massachusetts, 02115}
\vskip 1cm
A new link invariant is derived using the exactly solvable chiral Potts
model
and a generalized Gaussian summation identity.  Starting from
a general formulation  of
link invariants using edge-interaction spin models, we establish
the uniqueness of
 the invariant for self-dual models. We next apply
the formulation to the self-dual chiral Potts model, and obtain a
link invariant
in the form  of  a lattice sum defined by a matrix associated with
the link diagram.  A generalized Gaussian
summation identity is then used to carry out this lattice sum,
enabling us to cast  the invariant into a tractable form.
The resulting expression for the link
invariant is characterized by roots of unity and does not
appear to belong to the usual quantum group family of
invariants.
A
table of invariants for links with up to 8 crossings
is given.

\bigskip
\noindent
{\bf KEY WORDS}: Link invariants; chiral Potts model;
generalized Gaussian
summation identity.

\endpage

\noindent
{\bf 1. Introduction}

Knots and links are
embeddings of circles in
${\bf R}^3$
described by their projections onto a plane.
As projections  change their configurations
when the embedded circles are deformed
in ${\bf R}^3$, it is pertinent to ask
what is being preserved in the process of deformation.
Obviously, what is being preserved is intrinsic to the
topology of the link, and this leads to the consideration of
link invariants.

Link invariants are
algebraic quantities associated with planar projections,
which
remain unchanged when the links are deformed.
An exciting recent development in the theory of knots is the realization that
link invariants can be obtained from exactly solvable models in statistical
mechanics
(for reviews of this development see [1 - 3]).
Indeed, using solvable models in two dimensions,
it has been possible
to generate all known link invariants of the quantum group family,
including the Jones [4] and the Homfly [5]
polynomials.   More recent development on spin models and link invariants
can be found in [6 - 11].

In a recent {\it letter} [12], we reported a new link invariant
derived
from the solvable chiral Potts model [13, 14] and evaluated using
a generalized Gaussian summation identity [15].
The new invariant, which was earlier defined by Kobayashi {\it et al.}
[16]   without explicit evaluations, is
in the form of a polynomial of roots of unity.
In view of its novelty and the
 intimate relation with the Gaussian summation identity,
it seems useful to
provide  details of our analysis. This is the purpose of the present
paper.  In addition,
we also present  a general  duality consideration, a discussion
of some properties and relations of
our invariant to other known polynomial invariants, and
 a  table of the invariant
for links with 8 or less crossings.

This paper is organized in the following manner. In section 2 we recall the
formulation of link invariants using edge-interaction
models, paying particular attention
to models with chiral interactions.
In section 3 we present the formulation of a duality
relation for general edge-interaction spin models and show
that it leaves unchanged
the invariant derived from self-dual models.
In section 4 we
recall the solvable
chiral Potts model, and a certain infinite rapidity limit
of its vertex weights is introduced in section 5.
Section 6 deals with the detailed evaluation of invariants.
We derive the Skein relation satisfied by the invariant and
discuss relations with other known invariants in section 7.
Several technical points and some properties of the invariant are discussed
in section 8.
In particular, we establish the identity of the invariant
with its mutant, a result we extend to any invariant
derivable from edge-interaction spin model.
In Appendix A we describe the
summation identities known as the Gaussian summation formula.
  A table
of invariants for all links with up to $8$
crossings is given in Appendix B.
\medskip
\noindent {\bf 2. Link invariants from edge-interaction models}

Link invariants can be generated from solvable two-dimensional models
in statistical mechanics [1 - 3].
Here, we briefly review
the formulation involving edge-interaction models.

Starting from a link $K$ which we assume to be oriented, one constructs
a directed lattice ${\cal L}$ by regarding lines of the link as
lattice edges and line crossings as lattice vertices.  This leads
naturally to two types of vertices, $+$ and $-$, corresponding to
the two kinds of line crossings $+$ and $-$ in the link
as shown in Fig. 1.  Now the lines
divide faces of ${\cal L}$ into two
subsets, with one subset
neighboring only faces of the other subset, and vice
versa.
This permits
us to introduce a  spin model with spins residing in one subset of
the faces and interactions extending across
the line intersections.
The spins and the interactions form a line graph
$G$. To help one to visualize, it is customary to
shade the faces in which the spins reside [17].
Then, the line graphs for the
two different shadings are mutually
dual.  The example of the two different face shadings for the link $8^2_{15}$
is shown in Fig. 2.  Note that
there exist four distinct types
of line crossings
and, consequently, four  types of spin interactions.
These four different configurations,
shown in Fig. 1, possess  weights
$$ u_{\pm}(a-b), \hskip 1cm {\overline u}_{\pm}(a-b), \eqno(1)$$
where $a , b = 1,2, \cdots, N$ are the spin states
of the two interacting spins.
Here, we have assumed quite generally that
the interactions can be chiral, namely,
$u_\pm (a) $ can be different from $u_\pm(-a)$.
The example of the two line graphs for the link
$8^2_{15}$ is shown in Fig. 2 with the type of interactions
explicitly noted.

Let $Z(u_{\pm}, \ub_\pm)$ be the partition function of the spin model
for a given face shading.
Then, the formulation of link invariants using spin models [1, 3]
dictates that
$Z(u_{\pm}, \ub_\pm)$ is
a link
invariant, provided that the
Boltzmann weights satisfy certain conditions imposed by Reidemeister
moves [18].  Reidemeister  moves
 are elementary moves of lines in the knot
projection when links are deformed in ${\bf R}^3$.
The possible Reidemeister moves that can occur
are those shown in Fig. 3, where for each line movement
we must allow the  two possible kinds of
face shadings.
The desired Reidemeister conditions can be
read off  from the figure, leading to
$${\ub}_\pm (0)=1 \eqno(2a)$$
$${1\over {\sqrt N}}\sum _{b=0}^{N-1} u_\pm(a-b)=1 \eqno(2b)$$
$$u_+(a-b)u_-(a-b)=1 \eqno(2c)$$
$${1\over N} \sum _{b=0}^{N-1} {\overline u}_+ (a-b) {\overline u}_-(b-c)
= \delta_{ac}
\eqno(2d)$$
$${\overline u}_+(a-b) {\overline u}_-(b-a) =1 \eqno(2e)$$
$${1\over N} \sum _{b=0}^{N-1} u_+ (a-b) {u}_-(c-b) = \delta_{ac}
\eqno(2f)$$
$$ {1\over \sqrt N} \sum_{d=0}^{N-1}
u_-(a-d) \ub_-(b-d) \ub_+(d-c)
= \ub_-(b-c)u_-(a-c)u_+(a-b) \eqno(2g)$$
Provided that conditions (2a) - (2f) are met, the quantity
$$I_K(N)\equiv N^{(c_D-S-2)/2} Z(u_\pm,   \ub_\pm), \eqno(3) $$
where $S$ is the number of spins (shaded faces) in ${\cal L}$
and $c_D$ is the number of connected components of the line graph associated
with the other (dual) choice of face shading,
is an invariant for the link $K$ [3, 12].
Note that $c_D=1$ for connected ${\cal L}$.
Note also that
the normalization of (3)  has been taken to be
$I_{\rm unknot} (K)= 1$.
\medskip
\noindent{\bf 3. Duality relation for spin models}

In this section we present the formulation
of a general duality relation for two-dimensional
edge-interaction spin models in a form suitable for our consideration.

Consider an $N$-state spin model on a line graph $G$ with
$S$ sites.  The spins interact with a
generally chiral interaction
$u_{ij}(a-b)$ along the lattice edge $\{i, j\}$ connecting
sites $i$ and $j$ in respective
spin states $a$ and $b$.
Consider the dual model whose spins are on the dual graph $G_D$,
with $S_D$ sites and with interactions
$$u^{(D)}(n) = {1\over \sqrt N} \sum _{m=1}^N \o^{mn}u(m)  \eqno (4)$$
whose inverse is
$$u(n) = {1\over \sqrt N} \sum_{m=1}^N \o^{-mn}u^{(D)}(m), \eqno(5)$$
where $\o=e^{2\pi i/N}$.

Let $c$ and $c_D$ be the respective numbers of connected components
of $G$ and ${G}_D$. (For connected lattices
we have $c=c_D=1$).  Let $Z(\{u\})$ and $Z^{(D)}[\{u^{(D)}\}]$
be the respective partition functions of the spin models on $G$
and ${G}^D$.  Then, following Wu and Wang [19] one establishes the
identity
$$ Z(\{u\}) = N^{c-S_D}Z^{(D)}[\{\sqrt N u^{(D)}\}], \eqno(6)$$
where the partition function on the right-hand side of (6) is
defined with interactions $\sqrt N u^{(D)}$ in place of $u^{(D)}$.
Using the Euler relation generalized to  disjoint
graphs
$$ S+S_D= c+c_D +E, \eqno(7)$$
where $E$ is the number of edges (which is the same for ${G}$
and ${G}_D$), one obtains
$$ N^{(c_D-S-2)/2} Z(\{u\}) = N^{(c-S_D-2)/2} Z^{(D)}[\{u^{(D)}\}]. \eqno(8)$$
This is the desired duality relation.

For chiral interactions for which lattice edges are directed, the
orientation of a lattice edge on ${G}_D$ is that of the
corresponding lattice edge on ${G}$ rotated $90^\circ$
[19].  Furthermore, the shading of the other set of faces,
namely, choosing  spins to reside in the other set of faces,
corresponds
to the interchange of $u_\pm \leftrightarrow \overline u_\pm$
(Cf. Fig. 1).
Therefore, a $\overline u$ interaction on $G$ corresponds
in the dual space to a $u_\pm^{(D)}$
interaction, and a
$u$ interaction on $G$ corresponds
on $G_D$ to a $\overline u_\pm^{(D)}$.  This leads to
the duality relation
$$ N^{(c_D-S-2)/2} Z(u_\pm, \overline u_\pm)
= N^{(c-S_D-2)/2} Z^{(D)}
[\overline u_\pm^{(D)},u_\pm^{(D)} ]. \eqno(9)$$
It follows that for self-dual models satisfying
$ \overline u^{(D)}_\pm
=u_\pm, \>\> u^{(D)}_\pm = \overline u_\pm,$
the invariant (3) evaluated using either scheme of
face shading is identically the same.  This
conclusion applies to both the chiral
and non-chiral self-dual models.
\medskip
\noindent{\bf 4. The chiral Potts model}

The $N$-state chiral Potts model  is a spin model
whose  Boltzmann weights $W(n)$ and ${\overline W}(n)$
are  chiral and $N$-periodic. Namely,
quite generally we have
$W(-n)\not=W(n), {\overline W}(-n) \not={\overline W}
(n)$ and the equalities
$W(n)=W(n+N)$,  ${\overline W}(n)
={\overline W}(n+N)$.
An important recent advance in two-dimensional
lattice statistics has been the discovery of
an exact integrable manifold of the chiral Potts model [13, 14].
In the notation of [14],
the integrable vertex weights  are
best described by introducing rapidities ${a_p, b_p, c_p, d_p}$ to
auxiliary lines drawn as shown in Fig. 4.  The
rapidities satisfy the $N$-periodicity condition
$${{a_p^N \pm b_p^N} \over { c_p^N\pm d_p^N}} =\lambda_\pm, \hskip .8cm
{\rm independent\>\>\> of \>\> p}.\eqno(10)$$
Then the weights can be written , for $0\leq n \leq N-1$, as
$$\eqalign {g_{pq}(n)\equiv &
{{W_{pq}(n)}\over{W_{pq}(0)}} =
  \prod _{j=1}^n {{d_pb_q-a_pc_q\o^j}\over
  {b_pd_q-c_pa_q\o^j}}
    \cr
  {\overline g}_{pq}(n)\equiv &
{{{\overline W}_{pq}(n)}\over {{\overline W}_{pq}(0)}} =
    \prod_{j=1}^n {{\o a_pd_q-d_pa_q \o^j} \over
    {c_pb_q -b_pc_q \o^j}}.\cr } \eqno(11)$$

The weights (11)
satisfy the relations
$$g_{pp}(n)=1, \hskip 1cm \gb_{pp}(n) = \delta_{n0}\eqno(12a)$$
$$g_{pq}(0)=\gb_{pq} (0) =1 \eqno(12b)$$
$$g_{pq}(a-b) g_{qp}(a-b) =1 \eqno(12c)$$
and the Yang-Baxter equation
$$
\sum_{d=0}^{N-1} W_{pr}(a-d){\overline W}_{qr}(b-d)
\overline W_{pq}(d-c) = R_{pqr}
\overline W_{pr}(b-c)W_{qr}(a-c)W_{pq}(a-b) \eqno(12d)$$
where
$$ \eqalign { R_{pqr}&=f_{pq}f_{qr}/ {f_{pr}} \cr
 f_{pq} &= \biggl[ \prod_{m=0}^{N-1}
 {{\sum_{k=1}^N \o^{km}{\overline W}_{pq}(k)} \over {W_{pq}(m) }}\biggr]
 ^{1/N}.\cr} \eqno(13)$$
For
$$\eqalign { c_p &=d_p=1 \cr
             a_p^N+b_p^N& = I = {\rm constant}\cr} \eqno(14)$$
the model is self-dual [20] and $f_{pq}$ is independent of $p$ and $q$.

\endpage
\noindent{\bf 5. The  infinite rapidity limit}

To generate link invariants we need first to
identify the weights
$u_\pm$ and $\ub_\pm$ appearing in (2a) - (2f).
Comparing (2) with (12) one observes that an obvious choice is
 $\{u_+, u_-, \overline u_+, \overline u_-\} \sim
 \{g_{pq}, g_{qp}, \gb_{pq}, \gb_{qp}\}$ in (14).
This leads us to introduce
the infinite rapidity limits
$$\eqalign{ u_{+}(n)=&
A_{+}\lim _{b_p\rightarrow \infty}g_{pq}(n),\hskip 0.5cm
        u_{-}(n)= A_{-}\lim _{b_q\rightarrow \infty}g_{pq}(n)\cr
{\overline u}_{+}(n)=&
B_{+} \lim _{b_p\rightarrow \infty}{\bar g}_{pq}(n),\hskip0.5cm
 {\overline u}_{-}(n)=B_{-} \lim _{b_q\rightarrow \infty}{\bar g}_{pq}(n),\cr}
\eqno(15)$$
where $A_\pm$ and $B_\pm$ are constants  to be determined.
It turns out  that conditions (2a) - (2g) can be satisfied only for
the self-dual model (10) and, within this framework,
it can be seen that  there is no loss of generality
to take $I=0$ in (14). This leads us to write
$$ a_p=\o^{\ell- 1/2} b_p, \eqno(16)$$
where $\ell$ is an integer, and
$$\eqalign{u_{\pm}(n)=& A_{\pm}(-1)^{n}\o^{\pm (n\ell+ {n^2/2})},\cr
{\overline u}_{\pm}(n)=& B_{\pm}(-1)^{n}\o^{n\ell \mp {n^2/2}}.\cr}
\eqno(17)$$

{}From (2a) and using the Gaussian summation identity (A2) with $M=1$ in
Appendix A to evaluate (2b), one obtains
$$ \eqalign {A_{\pm}&=(-1)^\ell\o^{\pm \ell^2/2}e^{\pm i\pi(N-1)/4} \cr
 B_{\pm}&= 1.\cr}\eqno(18)$$
It can then be verified that conditions (2c) - (2g) are now all satisfied
by (17) and (18).
It can also be  verified  from (4)
that we have
$$\eqalign{ u_\pm^{(D)}(n) &= \overline u_\pm(-n) \cr
           \overline u_\pm^{(D)}(n)&= u_\pm(n).\cr} \eqno(19)$$
Thus, as discussed at the end of section 3,
the invariant $I_K(N)$ is independent of the face shading chosen.
This is the desired result.

The invariant
 $I_K(N)$ is also independent of $\ell$ in (16).  To see this,
we observe from the second line of (17) that the $\overline u_\pm(a-b)$
interaction between spin states $a$
and $b$ contributes to the summand of the partition function
an $\ell$-dependent factor
$\o^{\ell a}\o^{-\ell b}$.  The two    factors in this product can be split
and associated separately to the two
corners of  the two shaded polygons meeting at the
 line crossing (Cf. the
second row of Fig. 1).  Next we collect these factors for each polygon.
Let the spin state of a shaded polygon
be  $a$.  Then,
to each
corner of the polygon with two outgoing (incoming) arrows
[source (sink) of arrows]
we have a factor
 $\o^{-\ell a}$ ($\o^{\ell a}$).
As demonstrated in the example
shown in Fig. 5, there is always an equal  number of
sources and sinks around a polygon.
Thus, the $\ell$-dependent factors around each polygon cancel
out and, as a result,
 the link invariant $I_K(N)$ does not change its value
if we  set
$\ell =0$ in $\overline u_\pm$.
As discussed in section 7 below, every link has a
special  projection in
which all crossings are of the type $\overline u_\pm$.
It follows that
$I_K(N)$ is independent of $\ell$.
For self-dual models the dual of the
$u_\pm$ weights are the $\overline u_\pm$ weights.
Therefore, it also follows that we  can set
$\ell =0$ in $u_\pm$ (when evaluated in the dual space), and
hence deduce directly that $I_K(N)$ is independent of $\ell$.

Finally,
combining (17) with (18) and $\ell=0$,
we arrive at the expression
$$\eqalign {u_\pm(n) = &
(-1)^{n} e^{ \pm i(N-1)\pi/4} \o^{\pm n^2/2}\cr
   \ub_\pm(n)  = &(-1)^n \o^{ \mp n^2/2}.\cr} \eqno(20)$$
The substitution of (20) into (3) leads to the desired link
invariants.
\medskip
\noindent {\bf 6. Evaluation of knot invariants}

The link invariant $I_K(N)$ for a link $K$ is evaluated by substituting
(20) into (3). Noting that the
essential difference between the weights
$u$ and $\overline u$ in (20) is the sign of the factor
  $n^2/2$ in the exponent, it is useful to
explicitly note this sign, $+$ or $-$, on the line graph,
and this leads us to consider the signed graph associated with a face
shading.  Specifically, to each edge $\ell$ in a line graph $G$, we
assign a number $\epsilon_{\ell} = \pm 1$
according to the following rules: if the associated line crossing
is of type $\overline{u}_{+}$ or $u_{-}$, assign $\epsilon_{\ell}=+1$,
if the crossing is of type
$\overline{u}_{-}$ or $u_{+}$, assign $\epsilon_{\ell}=-1$.
The example of the two signed graphs associated with the line graphs for
the link $8_{15}^2$ of Fig. 1 is shown in Fig. 6.  Note that this sign
is determined by the face shading only, independent
of the orientation of lines in ${\cal L}$.
Thus, our choice of the
sign agrees with that of Fig. 48 of Ref.
[3].

To facilitate
bookkeeping, we now introduce an $S\times S$ (incidence) matrix
${\bf Q}$ with elements
$$\eqalign {{Q}_{ij} =& \sum_{\ell=<i,j>}
\epsilon _{\ell}, \hskip 1.4cm i\not= j \cr
Q_{ii}=& -\sum_{k (\neq i)} Q_{ik},\cr }\eqno(21)$$
where the summation in the first line is over {\it all} edges $\ell$
connecting the $i$-th and $j$-th sites.
The matrix ${\bf Q}$ is obviously symmetric; it also has the property that
the sum of each row
or column vanishes.  Matrices possessing these properties are singular,
all cofactors  are equal,
and  the cofactors
generate spanning trees of the graph $G$
[21].

Let ${\bf n} =(n_1,...,n_{S})$
be an integer-valued
vector whose component $n_i=0, 1,...,N-1$, $i=1,...,S$, denotes the
state of the $i$th spin on $G$.  Further introduce a vector ${\bf z}$ with
components
 $z_{i} =  Q_{ii}/2, i=1,...,S$.
Then the
link invariant (3), after the substitution with (20), assumes the form
$$
I_{K}(N) = N^{(c_D-S-2)/2 } e^{\pi i  (N -1) \zeta(K)/4}
\sum_{n_1,\cdots,n_{S}=0}^{N-1}
{\rm exp}\biggl[{\pi i \over N} \u{n}\cdot ({{\bf Q}} \u{n}) +
2 \pi i \u{n}\cdot \u{z} \biggr] \eqno(22)$$
where
$\zeta(K)$  is
the number of
$u_+$ weights minus the number of
$u_-$ weights in $K$.

Due to the $N$-periodicity of Boltzmann weights, there is no loss of generality
to fix one of the $S$ spins, say, the $S$th, in the spin state $n_S=0$.
Then the summation over $n_S$ in (21) gives rise to a factor $N$ and one
obtains
$$
I_{K}(N) = N^{(c_D-S)/2 } e^{\pi i  (N -1) \zeta(K)/4}
\sum_{n_1,\cdots,n_{S-1}=0}^{N-1}
{\rm exp}\biggl[{\pi i \over N} \u{n}\cdot ({\bf M} \u{n}) +
2 \pi i \u{n}\cdot \u{y} \biggr], \eqno(23)$$
where
$\u{n}$ and $\u{y}$ are  $(S-1)$-dimensional vectors,
${\bf M}$ is the $(S-1)\times(S-1)$ cofactor matrix
of ${\bf Q}$ obtained by deleting the $S$-th row and column,
and
the summation is ($S-1$)-fold.
Note that we have always
$$N(2y_i+M_{ii}) = 2NM_{ii}= {\rm an \>\> even\>\> integer.} \eqno(24)$$

The expression (23) subject to (24)
can be evaluated using a generalized Gaussian
summation formula   given in Appendix A.  Provided that the matrix
${\bf M}$ is nonsingular, using (A5) we find
$$
\eqalign{I_{K} (N)= N^{({c_D}-1)/2}& e^{\pi i  (N -1) \zeta(K)/4}
e^{\pi i  \eta(\u{M})/4}\cr
&\times{1\over {\sqrt D}}\sum_{\u{n} \in \Delta}
{\rm exp}\biggl[-\pi i N(\u{n} + \u{y})\cdot
[\u{M}^{-1}
(\u{n} + \u{y})]\biggr],\cr}  \eqno (25)$$
where  $\u{n}=(n_1,...,n_{S-1})$, ${\bf y}=(y_1,...,y_{S-1}),
y_i=M_{ii}/2$, $D=|{\rm det} {\bf M}|$,
$\Delta$ is the fundamental domain (unit cell) of the
lattice formed by the collection of vectors $\u{Mn}$,
and $
\eta(\u{M}) $ is the signature of {\bf M}, namely, the
number of positive eigenvalues
minus the number of negative eigenvalues.
Note that $N$ now appears as a parameter, instead of a summation
limit, in (25).
The expression (25) completes the evaluation of $I_K(N)$.

It is instructive to illustrate the evaluation of (25) for the link
$8^2_{15}$ shown in Fig. 1.  The two signed graphs corresponding to the
two different face shadings are shown in Fig. 6, from which one
obtains the matrices
$${\bf M} =\pmatrix{-2&-1&1&0&0\cr  -1&2&0&-1&0 \cr
   1&0&-2&1&0\cr 0&-1&1&-1&1\cr 0&0&0&1&-2\cr},
\hskip 1cm \pmatrix {1&2&-2\cr 2&0&-2\cr -2&-2&5\cr} \eqno(26)$$
for the two shadings, respectively.  In the first case one finds further
$\zeta(K)=-2, \eta({\bf M})= -3, D=8, $
$\Delta = (-2,1,0,0,-1), (-1,0,-1,1,-1),(-1,0,0,0,0), $
\nextline
$(-1,1,0,0,-1), (-1,1,-1,0,0), (0,0,-1,1,-1), (0,0,0,0,0),$
and $(0,1,-1,0,0)$.
In the second case ${\bf M}$ is a $3\times 3$
matrix.  One finds
$\zeta(K)=2, \eta({\bf M})=-3, D=8$,
$\Delta = (-2,-1,2),
(-1,-1,1), (-1,0,0),(-1,0,1), (0,0,-1), (0,0,0), $
$(0,1,-2),$
and $(1,1,-3).$
In either case, (25) yields the invariant
$$I_K(N) = {1\over {\sqrt 2}}e^{-i\pi/4}\biggl(1+e^{-7\pi iN/8} +e^{-3\pi
iN/2} +e^{-15\pi iN/8}\biggr).$$

We have used a computer program to compute invariants from (25).
Generally, for a given link $K$ with a given orientation, the matrix ${\bf
M}$ and the number $\zeta(K)$ can be read off from the link
diagram and are used as inputs.
 The computer then searches for all sites
in the fundamental domain $\Delta$ and evaluates
(25) term by term.
We include in Appendix B a table of invariants for all links with
8 or less crossings.
We note that our invariant assumes the same form for some links.
Examples are the pairs
$ \{6_2,7_2\},\{6_3, 8_1\}, \{7_5,8_2\},\{8_6,8_7\},
\{8_{10}, 8_{11}\}, \{8_{12}, 8_{13}\}$,
and $\{7_7^2, w(K)=-1; 4_1^2, w(K)=4\}$.
\medskip
\noindent
{\bf 7. Relations to other invariants and Skein relations}

The matrix {\bf M} occurring in (22) has been used
in previous studies of link invariants, in particular by
Goeritz [22], and has come to be known as the Goeritz matrix
[23]. Traldi [24] introduced an extension of
{\bf M}, which he called the modified Goeritz matrix, and
this was the starting point for Kobayashi {\it et al.}
[16] who built their invariant $T_{N}(K)$
from this matrix. As alluded to in section 1, $T_N(K)$
 turns out to be identical to $I_{K}(N)$, although  in
a different form and without an explicit evaluation.
The Goeritz matrix is also related to the Seifert matrix
of the link.  For completeness,
we briefly define relevant notions
 and describe some
related results.

Consider an oriented link diagram with a shading, where
the infinite region is one of the shaded regions. Denote by
${\alpha}_{0}, {\alpha}_{1}, \dots, {\alpha}_{p}$
these shaded
regions, where $p=S-1$ and ${\alpha}_{0}$ is the infinite region.
Denote by ${\beta}_{1}, \dots, {\beta}_{q}$ the regions in the
dual shading. The diagram is a {\it special projection}
if the following conditions are satisfied:

i) all $\beta$-regions have consistently oriented
boundaries.

ii) all $\beta$-regions are topological disks, without
holes.

It is not difficult to show that every link has a special projection
[23]. Since the link diagram is the projection of
a link in space onto a plane,
we may consider the $\beta$-regions in a special
projection as the projection onto the plane
of an oriented surface in space, whose boundary
is the link. This surface is  the Seifert surface
of the link. It follows that $p = 2g$,
where $g$ is the genus of the  Seifert surface.

Suppose we have a special projection of a link.  Place
spins in the (shaded) $\alpha$-regions
as described in section 2 and define the {\bf Q} matrix
associated with the projection as in section 6.
Omitting the infinite region
${\alpha}_{0}$ reduces {\bf Q}  to the matrix {\bf M}.
Since each $\alpha$-region has an even number of boundary
components, the diagonal entries of {\bf M} are always even. Furthermore,
all crossings are now of type $\overline u_{\pm}$, so
$\zeta(K) = 0$.
Hence the invariant $I_K(N)$ given by (25) reduces to the simple
form
$$
I_{K}(N) = N^{(c_D-1)/2} e^{\pi i \eta({\bf M})/4} {1 \over \sqrt{D}}
\sum_{n \in \Delta} {\rm exp}\biggl[-\pi i N {\bf n} \cdot {\bf M}^{-1}
{\bf n}\biggr]. \eqno (27)
$$
In this situation $\eta({\bf M})$ is itself a link invariant, and is
called the {\it signature} of the link. As noted in [12],
$D = |\Delta(-1)|$, where $\Delta(t)$ is the
Alexander polynomial. This fact can also be seen
by considering a Seifert matrix as  follows.

Let $a_{i}$ be a simple closed curve on the Seifert surface
which projects to a simple closed curve which encloses
the region $\alpha_{i}$, in the positive direction, but
does not enclose any other $\alpha$-regions.
Let $a_{i}^{-}$ be the ``push-off", or
a small displacement, of $a_{i}$ from the surface
in the direction opposite to the orientation of
the surface. Then  the Seifert matrix {\bf S} is a
$p\times p$ matrix whose $(i, j)$-th element  is
the linking number of $a_{i}^{-}$ and $a_{j}$ for all
$1 \leq i,j \leq p$ [23]. It is a straightforward calculation to show that
${\bf M} = {\bf S} + {\bf S}^T$, where ${\bf S}^T$ is the
transpose of ${\bf S}$.
It is also known that the Alexander polynomial is
$\Delta (t) ={\rm det} ({\bf S}^T- t{\bf S})$ and that the
signature of the link is
that of the matrix
${\bf S} + {\bf S}^T$.
These results imply $D=|{\rm det} {\bf M}| = |\Delta (-1)|$.

Since the matrix {\bf M} in (23) can be expressed in terms of the
Seifert matrix,
it is natural to ask whether
$I_{K}(N)$ contains different information than the
Alexander polynomial $\Delta (t)$ and the signature invariant,
which are also expressed in terms of the Seifert matrix.
One example shows that the answer is yes.
Let $\overline K$ denote the mirror image of $K$.
Then, it can be verified that
the four
{\it nonequivalent} knots $6_{1}$, $\overline{6_{1}}$, $9_{46}$
and $\overline{9_{46}}$ share the same Alexander
polynomial, and signature.  However,
the invariant $I_{K}(N)$ distinguishes
between $6_{1}$, $\overline{6_{1}}$ and $9_{46}$, although
not  between $9_{46}$ and $\overline{9_{46}}$.
On the other hand, the Alexander polynomial distinguishes the links
$\{8_{12}, 8_{13}\}$ for which $I_K(N)$ is identical.

Finally, the invariants $I_{K}(N)$
satisfy a Skein relation [12].
In particular, $I_K(2)$
and $I_{K}(3)$ are related
to the Jones polynomial  at special values.
The Jones polynomial $V_K(t)$ is determined by the
Skein relation
$$
{1 \over t} V_{K_{+}}(t) - t V_{K_{-}}(t) =
(\sqrt{t} - {1 \over \sqrt{t}}) V_{K_{0}}(t), \eqno(28)
$$
where the links $K_{+}$, $K_{-}$ and $K_{0}$ differ only at one
crossing, as shown in Fig. 7.  For definiteness we choose the shading
shown in Fig. 7. Then the corresponding partition functions
$Z_{+}$, $Z_{-}$ and $Z_{0}$ differ only by the factor
arising at this crossing, which is respectively
$u_{+}(n)$, $u_{-}(n)$ and $1$,
where $u_\pm (n)$ is given in (20). It is easy to show that the following
identities hold for $N=2,3$:
$$
\eqalign
{&N=2: \qquad u_{+}(n) + u_{-}(n) = \sqrt{2}, \hskip 2cm n=0,1 \cr
&N=3: \qquad e^{-\pi i /6} u_{+}(n) +  e^{\pi i /6} u_{-}(n) =
1, \quad n=0,1,2. \cr} \eqno(29)
$$
These imply the same relations for $Z_{+}$, $Z_{-}$ and $Z_{0}$.
Comparing with the Skein relation (28) for the Jones polynomial
we see that
$$\eqalign {
I_{K}(2)& = (-1)^{c(K) + 1} V_{K}(-i) \cr
I_{K}(3)& = (-1)^{c(K) + 1} V_{K}(e^{-\pi i /3}), \cr } \eqno(30)$$
where $c(K)$ is the number of components of the link $K$.
The invariants (30) can further be related to the Homfly polynomial
$P_K(t, z)$ by using the identity $(-1)^{c(K)+1} V_K(t)
=P_K(-t, \sqrt t - 1/\sqrt t)$.

As discussed in [12], the
invariants for higher values of $N$ also satisfy certain Skein
relations, but with higher order crossings.
For example,
the invariant $I_{K}(4)$ satisfies
$$
I_{K_{2+}}(4) - I_{K_{+}}(4) - i I_{K_{-}}(4) =  -i I_{K_{0}}(4)
\eqno(31)
$$
where $2+$ represents a crossing with 2 consecutive twists of the type
$u_+$ shown in Fig. 7.
Generally,
the invariant $I_K(N)$ satisfies a Skein relation
connecting
$I_{K_0}(N),I_{K_-}(N), I_{K_+}(N), \cdots, I_{K_{[N/2]+}}(N),$
where $[N/2] = N/2$ for $N={\rm even}$ and $[N/2] =(N-1)/2$ for $N={\rm odd}$,
and $n+$ is a crossing
of $n$ consecutive twists of the type $u_+$.
The Skein relation is
obtained by writing out the identity
$$ u_-(n)\prod_{p=0}^{[N/2]}\biggl[u_+(n)-u_+(p)\biggr] =0,
\hskip 0.8cm n=0,1,...,N-1,\eqno(32)$$
and making use of (2c). These relations,
which are reminiscent of the
Skein relations satisfied by the Akutsu-Wadati polynomials [25, 3],
are not
very useful for evaluating the invariant.
\endpage
\noindent {\bf 8. Discussions}

\noindent
{\it Mirror image and orientation reversals}:

The invariant for
the mirror image of a link is obtained by taking the complex
conjugation.
This follows from the fact that, in a mirror image, one
interchanges
$u_+ \leftrightarrow u_-$, $\overline u_+ \leftrightarrow
\overline u_-$, and hence ${\bf M} \leftrightarrow -{\bf M}$,
and by inspection of (20) and (23) one finds that these changes induce
only a complex
conjugation.  Also,
since ${\bf M}$
is independent of line orientations,
from (23) we see that  the reversal of the orientation of individual
components in a link introduces only an overall factor $e^{i(N-1)\pi
\Delta\zeta/4}$, where $\Delta \zeta$ is the induced change of $\zeta(K)$.

\noindent
{\it Invariants for split links}:

In standard notation, a link $K$ is split if it can be
deformed so that
a hyperplane in ${\bf R}^3$ separates the link into two disjoint nonempty
pieces, $K_1$ and $K_2$, say.
By choosing the shading that leaves the infinite region unshaded
as shown in Fig. 8, it is
clear from a consideration of (3) that
$$I_K(N) = \sqrt N I_{K_1}(N) I_{K_2}(N) , \eqno(33)$$
where the factor of $\sqrt N$ comes from the changes of $c_D$.
It follows that $I_K(N)$ has a factor $N^{(m-1)/2}$,
where $m$ is the
number of disjoint pieces contained in the split link $K$.

\noindent
{\it Invariants for connected links}:

The connected sum of two links $K_1$ and $K_2$, the
link $K_1\#K_2$, is obtained by cutting open both links and joining
them as shown in Fig. 8.  Denote  by $K$ the disjoint union of links
$K_1$ and $K_2$ before they are connected.
By considering (3) with
the infinite region shaded and using (33), one finds
$$ I_{K_1\#K_2}(N) = N^{-1/2} I_K(N) =
I_{K_1}(N) I_{K_2}(N) .  \eqno(34)$$
Therefore,
like
the Jones and Homfly polynomials,
$I_K(N)$ factorizes over connected sums of links.

\noindent
{\it Mutant links}:

We next review the notion of mutant links [26],
and show that $I_{K}(N)$ is unchanged under this operation.
Suppose there is a simple closed curve in the plane
which cuts a link $K$ at four points only. By deforming the
projection, the interior of the curve
can be placed inside a box as shown in
Fig. 9(a), with the orientations of the four incoming and outgoing
lines as indicated. The part of the link inside the
box is called a tangle.
If the four lines are cut and reconnected after a half twist
is put on the incoming lines, and the {\it opposite} half-twist
on the outgoing lines as in Fig. 9(b),
we get a new
link $\tilde K$ which is called a mutant of $K$.

Suppose now that we calculate $I_{K}(N)$ by using (3). We choose the
shading so that spins are placed
in the regions on the left and right sides of the tangle.
The orientations  of incoming and outgoing lines of the tangle
mean that these are different regions.
The summand of the partition function for $\tilde K$ has
an extra factor $u_+(a-b)u_-(a-b)$, where $a$ and $b$ are the spins
in the left and right regimes. By (2c), which comes from
the Reidemeister
move, this factor is 1.
Hence the link $K$ and its mutant $\tilde K$ have the
same partition function, and the same invariant.

Note that the above argument is quite general and establishes
the general result that an invariant
and its mutant are identical,
provided that the invariant
is derivable from an edge-interaction
spin model.
In particular, this applies to the Jones polynomial.

\noindent
{\it Links with singular {\bf M}}:

In writing down (25) we have assumed
that the matrix ${\bf M}$ is nonsingular, \ie, $D\not= 0$.  Indeed,
we find this condition satisfied by all links with 8 or less crossings
except $8^3_{10}$ and $8^4_3$.
Now the expression ${\bf n\cdot (Mn)}$
in the exponent of the summand in (23) is that of a quadratic
form over integral domains of the $S-1$ variables $n_i$, and
the condition $D=0$ says that the quadratic form is
singular.  It can be shown [27]
that such singular quadratic forms
can always be written as regular (nonsingular) ones over
a lesser number of integral variables.
After changing into these new variables,
the summations in (23) over the $\ell$ missing variables
can be performed, yielding a factor
$N^{\ell/2}$ where $\ell$ is the degeneracy of the
zero eigenvalue of {\bf M}, and the remaining summations
can therefore be evaluated using the Gaussian summation formula.

This procedure can be illustrated in a special case as follows.
When $D=0$ we know that
the rows of the matrix ${\bf M}$ are not all linearly
independent. This means that
there exist integers $c_i, i=2,3,...,S-1$
such that
$$M_{1j}=-\sum_{i=2}^{S-1} c_i M_{ij}, \hskip 1cm j=1,2,...,S-1.\eqno(35)$$
Since ${\bf M}$ is symmetric we have also
$$M_{i1}=-\sum_{j=2}^{S-1}  M_{ij}c_j, \hskip 1cm i=1,2,...,S-1, \eqno(36)$$
and, after substituting (36) into (35) and setting $j=1$,
$$ M_{11} =\sum_{i, j=2}^{S-1} c_iM_{ij}c_j. \eqno(37)$$
It follows that we have
$$\eqalign {{\bf n\cdot (Mn) }&= \sum_{i, j =1}^{S-1} n_i M_{ij} n_j \cr
&= n_1M_{11}n_1 +\sum_{j=2}^{S-1}n_1M_{1j}n_j
+\sum_{i=2}^{S-1}n_iM_{1j}n_j + \sum _{i,j=2}^{S-1} n_i M_{ij}n_j \cr
&= \sum_{i,j=2}^{S-1} (c_in_1-n_i)M_{ij}(c_jn_1-n_j) ,\cr}\eqno(38)$$
where we have used (35), (36), and (37) in reaching the last step in (38).

Provided that all
$c_i$'s are integers, the expression (38) is now a
quadratic form over $S-2$ integral variables, effectively
crossing out the first row and first
column of the matrix ${\bf M}$.  If this reduced quadratic form is again
singular, one repeats the process (again assuming integral $c_i$'s)
until the resulting quadratic form is regular.
The expression (23) can then be evaluated
by applying  the Gaussian summation
identity.
\medskip
\centerline {\bf  Acknowledgements}

We would like to thank Helen Au-Yang for pointing out the
resemblance of properties
of the chiral Potts model  weights (12) with those of the
Reidemeister moves (2), which has led to this investigation.
We also thank Alex Suciu for helpful discussions.
Work by FYW and PP has been supported in part by National Science
Foundation grants DMR-9313648, INT-9207261, and a Northeastern
University Research and Scholarship Development Fund grant.

\endpage

\centerline {\bf Appendix A. The generalized Gaussian  Summation}

In 1808 Gauss [28] obtained a remarkable summation identity,
now known as the
Gaussian Summation formula, which reads
$${1\over \sqrt{N}} \sum_{n=0}^{N-1}e^{{{2\pi i}\over{N}}n^2}={e^{i\pi/4}\over
\sqrt2}\biggl[1+e^{-i\pi N/2}\biggr].\eqno({\rm A}1)$$
The Gaussian summation identity (A1) can be generalized
in a number of ways [15, 29, 30].  A simple generalization
is the identity
$$ {1\over \sqrt N}\sum_{n=0}^{N-1}
e^{\pi iMn^2/N+2\pi i ny}
= {1\over \sqrt M} e^{i\pi/4} \sum_{m=0}^{M-1}e^{-\pi iN(m+y)^2/M},
\eqno ({\rm A}2)$$
valid  for  integral  $M,N$  and
$N(2y+M)={\rm an\>\> even\>\> integer}$, which recovers (A1) upon taking
$M=1$.
A multidimensional summation generalization which we use in arriving
(25) is the following.

Let ${\bf M}$ be  a
non-singular $p \times p$ symmetric matrix with
integer entries (positive or negative). We denote
by ${\bf n} = (n_{1}, \dots, n_{p})$, where $n_i$'s are integers,
a vector in $\z$,
and by ${\bf M} {\bf n}$ the vector with components
$\sum_{j=1}^{p} M_{ij} n_{j}$. The collection of
vectors $\{{\bf M{n} :  n} \in \z \}$ forms a
p-dimensional sublattice of $\z$. Let
$\Delta $ be a fundamental domain (unit cell) of this sublattice.
Similarly we define ${\bf x} =(x_1, ..., x_p)$, where $x_i$'s are real,
a vector in the $p$-dimensional space $\r$.
For all ${\bf x}, {\bf y} \in \r$, define
$$ \eqalign
{u({\bf {x}, {y}}) &=  {1\over{N^{p/2} }} \>
{\rm exp}\biggl[{\pi i \over N} \u{x}\cdot(\u{M}\u {x}) +
2 \pi i {\bf {x}\cdot {\bf y}} \biggr]\cr
v({\bf x}) &= {1\over \sqrt D} \>
{\rm exp} \biggl[- \pi i N {\bf x}\cdot ({\bf M}^{-1} {\bf x})\biggr],\cr}
\eqno({\rm A}3)$$
where $D = | {\rm det}{\bf M} |$.
Let $\eta({\bf M})$ denote the signature
of $\u{M}$, which is the number of positive eigenvalues
minus the number of negative eigenvalues of ${\bf M}$.
Also let
$C_{N}$ be the set of $N^p$ discrete points
$C_N= \{ {\bf n} \in \z : 0 \leq n_{i} \leq N-1,
\> i =1,...,p \}$.

For ${\bf y}\in \r$ satisfying the condition
$$N(2 y_{i} + M_{ii})={\rm even\>\> integer}, \hskip 0.5cm
\>\>i= 1,..,p,  \eqno({\rm A}4)$$
we have the {\it Generalized Gaussian Summation} formula
$$
\sum_{{\bf n} \in C_{N}} u({\bf n}, {\bf y}) =
e^{i \pi\eta({\bf M})/4}
\sum_{{\bf m} \in {\bf \Delta}} v({\bf m} + {\bf y}). \eqno({\rm A}5)
$$
For $p=1$  (A5) reduces to (A2).

\endpage

\centerline {\bf Appendix B. Table of invariants
for links with 8 or less crossings}

The invariant for a link $K$ is generally given by the expression
$$ I_K(N) =  \biggl({{N^{\ell/2}}\over \sqrt {D_1}}
\biggr)e^{ik\pi/4}\sum_{n=0}^{2D_2-1}c(n)
e^{-i n\pi N/D_2} \eqno (B1)$$
where $\ell=$ the degeneracy of the zero eigenvalue of {\bf M},
which is
equal to zero
unless $D=0$ as in $8^3_{10}$ and $8^4_3$.
The following table gives the values of
$[k,D_1,D_2]\{c(n)_n\}$ with  nonzero $c(n)$  listed.
For example, the invariant for the link $7_1$ is
$$ [2,7,7]\{1_0,2_2,2_4,2_8\} \rightarrow
{{1}\over
\sqrt 7}e^{i2\pi /4} \biggl(1+2e^{-i2\pi N/7} +2e^{-i4\pi N/7}
       +2e^{-i8\pi N/
7}\biggr).$$
The table also lists  $D=|{\rm det}{\bf M}|$
and $w(K) = n_+ -n_-$, the number of $+$ crossings minus
the number of $-$ crossings,
 to specify the direction of  line orientations.
Generally, $D=\sum_{n=0}^{2D_2-1}c(n)$
($\ell = 0$), $D_1$ and $D_2$ are factors of $D$ and, in
a few cases,
$D_2=2D$.
\bigskip
\settabs\+
$3_1\quad\quad$&$\hfill 3\quad\quad$&
     $\hfill +3\quad\quad$&
$[-i,3,3]\{1_{0},2_{4}\}$\cr 
\+$K$& $D$ & $w(K)$ &$I_K(N)$\cr
\+$3_1$&$3$&  $+3$&$[-2,3,3]\{1_{0},2_{4}\}$\cr
\+$4_1$& $5$&    $0$&$[0,5,5]\{1_{0},2_{2},2_8\}$\cr
\+$5_1$& $5$&   $+5$&$[4,5,5]\{1_{0},2_{4},2_{6}\}$\cr
\+$5_2$& $7$&  $+5$&$[-2,7,7]\{1_{0},2_{6},2_{10},2_{12}\}$\cr
\+$6_1$&  $9$&   $+2$&$[0,9,9]\{3_{0},2_{4},2_{10},2_{16}\}$\cr
\+$6_2$& $11$&    $+2$&$[-2,11,11]\{1_0,2_4,2_{12},2_{14},2_{16},2_{20}\}$\cr
\+$6_3$&  $13$&    $0$&$[0,13,13]\{1_0,2_2,2_6,2_8,2_{18},2_{20},2_{24}\}$\cr
\+$2_1^2$&  $2$&   $+2$&$[1,2,2]\{1_0,1_1\}$\cr
\+$4_1^2$&  $4$&     $-4$&$[3,4,4]\{1_0,2_3,1_4\}$\cr
\+&&$+4$&$[-1,4,4]\{1_0,1_4,2_7\}$\cr
\+$5_1^2$&  $8$&    $+1$&$[-1,2,8]\{1_0,1_3,1_{11},1_{12}\}$\cr
\+$6_1^2$& $6$&    $-6$&$[-3,6,6]\{1_0,2_5,2_8,1_9\}$\cr
\+&&$+6$&$[-1,6,6]\{1_0,1_3,2_8,2_{11}\}$\cr
\+$6_2^2$&  $10$&  $+6$&$[-3,10,10]\{1_0,1_5,2_8,2_{12},2_{13},2_{17}\}$\cr
\+$6_3^2$& $12$&  $-6$&$[3,12,12]\{1_0,2_4,4_7,1_{12},2_{15},2_{16}\}$\cr
\+&&$+2$&$[-1,12,12]\{1_0,2_3,2_4,1_{12},2_{16},4_{19}\}$\cr
\+$6_1^3$& $12$& $-6$&$[2,12,12]\{1_0,6_4,3_{12},2_{16}\}$\cr
\+&&$+2$&$[-2,12,12]\{3_0,2_4,1_{12},6_{16}\}$\cr
\+$6_2^3$&  $16$&  $0$&$[0,4,4]\{2_0,3_2,3_6\}$\cr
\+$6_3^3$&  $4$&   $-2$&$[0,4,4]\{3_0,1_4\}$\cr
\+&&$+6$&$[4,4,4]\{1_0,3_4\}$\cr
\+$7_1$& $7$&   $+7$&$[2,7,7]\{1_0,2_2,2_4,2_8\}$\cr
\+$7_2$&  $11$&   $+7$&$[-2,11,11]\{1_0,2_4,2_{12},2_{14},2_{16},2_{20}\}$\cr
\+$7_3$&  $13$&
$+7$&$[4,13,13]\{1_0,2_4,2_{10},2_{12},2_{14},2_{16},2_{22}\}$\cr
\+$7_4$& $15$&     $+7$&$[-2,15,15]\{1_0,2_6,4_{14},2_{20},2_{24},4_{26}\}$\cr
\+$7_5$& $17$&   $+7$&$[4,17,17]\{1_0,2_6,2_{10},2_{12},2_{14},2_{20},2_{22},
                        2_{24},2_{28}\}$\cr
\+$7_6$& $19$&
$+3$&$[-2,19,19]\{1_0,2_4,2_6,2_{16},2_{20},2_{24},2_{26},2_{28},
                        2_{30},2_{36}\}$\cr
\+$7_7$& $21$&
$+1$&$[0,21,21]\{1_0,2_6,4_{10},2_{12},2_{24},2_{28},4_{34},4_{40}
                                               \}$\cr
\+$7_1^2$&   $14$&
$-3$&$[3,14,14]\{1_0,2_4,1_7,2_8,2_{11},2_{15},2_{16},2_{23}\}$\cr
\+&&$+1$&$[1,14,14]\{1_0,2_1,2_4,2_8,2_9,2_{16},1_{21},2_{25}\}$\cr
\+$7_2^2$& $18$&
$-3$&$[1,18,18]\{3_0,2_5,2_8,3_9,2_{17},2_{20},2_{29},2_{32}\}$\cr
\+&&$+1$&$[0,18,36]\{2_{1},2_{7},3_{9},2_{25},2_{31},2_{49},2_{55},3_{63}\}$\cr
\+$7_3^2$& $16$&
$+3$&$[-1,4,16]\{1_0,1_7,1_{15},1_{16},1_{23},2_{28},1_{31}\}$\cr
\+$7_4^2$& $16$&
$+3$&$[-3,4,16]\{1_0,1_5,1_{13},1_{16},2_{20},1_{21},1_{29}\}$\cr
\+$7_5^2$& $20$&
$-7$&$[3,20,20]\{1_0,4_7,2_8,2_{12},2_{15},1_{20},4_{23},2_{28},
                                     2_{32}\}$\cr
\+&&$+1$&$[-1,20,20]\{1_0,4_3,2_8,2_{12},1_{20},4_{27},2_{28}
                                                 ,2_{32},2_{35}\} $\cr
\+$7_6^2$&  $24$&  $+1$&$[-1,6,24]\{1_0,1_3,2_{11},1_{12},1_{27},2_{32},2_{35},
                          2_{44}\}$\cr
\+$7_7^2$&  $4$&   $-1$&$[-1,4,4]\{1_0,1_4,2_7\}$\cr
\+&&$+7$&$[-3,4,4]\{1_2,2_5,1_6\}$\cr
\+$7_8^2$&$8$&   $+3$&$[-1,2,8]\{1_0,1_7,1_{12},1_{15}\}$\cr
\+$7_1^3$&  $20$& $+1$&$[0,20,5]\{3_0,6_2,2_3,1_5,2_7,6_8\}$\cr
\+$8_1$& $13$&   $+4$&$[0,13,13]\{1_0,2_2,2_6,2_8,2_{18},2_{20},2_{24}\}$\cr
\+$8_2$& $17$&   $+4$&$[4,17,17]\{1_0,2_6,2_{10},2_{12},2_{14},2_{20},2_{22},
                            2_{24},2_{28}\}$\cr
\+$8_3$& $17$&  $0$&$[0,17,17]\{1_0,2_2,2_4,2_8,2_{16},2_{18},2_{26},2_{30},
                                                   2_{32}\}$\cr
\+$8_4$& $19$&
$0$&$[2,19,19]\{1_0,2_2,2_8,2_{10},2_{12},2_{14},2_{18},2_{22},
                                       2_{32},2_{34}\}$\cr
\+$8_5$&  $21$&
$+4$&$[4,21,21]\{1_0,2_6,2_{12},2_{14},4_{20},2_{24},4_{26},
                                                     4_{38}\}$\cr
\+$8_6$& $23$&
$+4$&$[-2,23,23]\{1_0,2_{10},2_{14},2_{20},2_{22},2_{28},2_{30}
                                ,2_{34},2_{38},2_{40},2_{42},2_{44}\}$\cr
\+$8_7$& $23$&
$+2$&$[-2,23,23]\{1_0,2_{10},2_{14},2_{20},2_{22},2_{28},2_{30}
                                ,2_{34},2_{38},2_{40},2_{42},2_{44}\}$\cr
\+$8_8$& $25$&
$+2$&$[2,25,50]\{2_{13},2_{17},5_{25},2_{33},2_{37},2_{53},2_{57},
                         2_{73},2_{77},2_{93},2_{97}\}$\cr
\+$8_9$& $25$&   $0$&$[0,25,25]\{5_0,2_2,2_8,2_{12},
2_{18},2_{22},2_{28},2_{32},2_{38},
                            2_{42},2_{48}\}$\cr
\+$8_{10}$& $27$&
$+2$&$[-2,27,27]\{3_0,2_4,2_{10},2_{16},2_{22},2_{28},2_{34},
                                   6_{36},2_{40},2_{46},2_{52}\}$\cr
\+$8_{11}$&  $27$&
$+4$&$[-2,27,27]\{3_0,2_4,2_{10},2_{16},2_{22},2_{28},2_{34},
                                   6_{36},2_{40},2_{46},2_{52}\}$\cr
\+$8_{12}$& $29$&
$0$&$[0,29,29]\{1_0,2_2,2_8,2_{10},2_{12},2_{14},2_{18},2_{26},
                        2_{32},2_{40},2_{44},$\cr
 \+&&&$2_{46},2_{48},2_{50},2_{56}\}$\cr
\+$8_{13}$& $29$&
$+2$&$[0,29,29]\{1_0,2_2,2_8,2_{10},2_{12},2_{14},2_{18},2_{26},
                        2_{32},2_{40},2_{44},$\cr
     \+&&&   $ 2_{46},2_{48},2_{50},2_{56}\}$\cr
\+$8_{14}$&  $31$&
$+4$&$[-2,31,31]\{1_0,2_6,2_{12},2_{22},2_{24},2_{26},2_{30},2_{34},
    2_{42},2_{44}, $\cr
\+&&&$ 2_{46},2_{48},2_{52},2_{54},2_{58},2_{60}\}$\cr
\+$8_{15}$&   $33$&
$+8$&$[4,33,33]\{1_0,2_6,4_{10},2_{18},2_{22},2_{24},4_{28},
                            2_{30},4_{40},4_{46},4_{52},2_{54}\}$\cr
\+$8_{16}$& $35$&
$+2$&$[-2,35,35]\{1_0,4_4,2_{14},4_{16},2_{30},4_{36},4_{44},
                       4_{46},2_{50},2_{56},2_{60},4_{64}\}$\cr
\+$8_{17}$& $37$&
$0$&$[0,37,37]\{1_0,2_2,2_6,2_8,2_{14},2_{18},2_{20},2_{22},
      2_{24},2_{32},$\cr
\+&&&$ 2_{42},2_{50},2_{52},2_{54},  2_{56},2_{60},2_{66},2_{68},2_{72}\}$\cr
\+$8_{18}$&   $45$&
$0$&$[0,45,15]\{1_0,8_2,8_8,4_{10},2_{12},2_{18},4_{20},8_{22},
                             8_{28}\}$\cr
\+$8_{19}$& $3$&    $+8$&$[2,3,3]\{1_0,2_2\}$\cr
\+$8_{20}$& $9$&   $+2$&$[0,9,9]\{3_0,2_2,2_8,2_{14}\}$\cr
\+$8_{21}$& $15$&
$+4$&$[-2,15,15]\{1_0,2_{10},2_{12},2_{18},4_{22},4_{28}\}$\cr
\+$8_1^2$& $8$&   $+8$&$[1,2,8]\{1_0,1_1,1_4,1_9\}$\cr
\+$8_2^2$& $16$&
$+8$&$[3,4,16]\{1_0,1_3,1_{11},2_{12},1_{16},1_{19},1_{27}\}$\cr
\+$8_3^2$& $22$&  $-4$&$[1,22,22]\{1_0,2_1,2_4,2_5,2_9,2_{12},2_{16},2_{20},
                       2_{25},1_{33},2_{36},2_{37}\}$\cr
\+&&$+8$&$[3,22,22]\{1_0,2_3,2_4,1_{11},2_{12},2_{15},2_{16},2_{20},
              2_{23},2_{27},2_{31},2_{36}\}$\cr
\+$8_4^2$& $24$&  $-4$&$[1,6,24]\{1_0,1_3,2_4,2_{16},2_{19},1_{27},1_{36},
                                       2_{43}\}$\cr
\+&&$+8$&$[3,6,24]\{1_0,2_7,1_{12},1_{15},2_{16},2_{28},2_{31},1_{39}\}$\cr
\+$8_5^2$&
$26$&$+4$&$[-3,26,26]\{1_0,2_5,2_8,1_{13},2_{20},2_{21},2_{24},2_{28},
                             2_{32},2_{33},$\cr
\+&&&$2_{37},2_{41},2_{44},2_{45}\}$\cr
\+$8_6^2$&  $20$&
$0$&$[1,20,20]\{1_0,4_1,2_4,4_9,2_{16},1_{20},2_{24},2_{25},2_{26}
                         \}$\cr
\+&&$+8$&$[-3,20,20]\{1_0,2_4,2_5,2_{16},1_{20},4_{21},2_{24},4_{29},2_{36}
                                           \}$\cr
\+$8_7^2$& $30$&   $0$&$[-1,30,30]\{1_0,4_{11},1_{15},2_{20},2_{24},2_{35},
                  2_{36},2_{39},4_{44},2_{51},4_{56},4_{59}\}$\cr
\+&&$+4$&$[-3,30,30]\{1_0,2_5,2_9,2_{20},2_{21},2_{24},4_{29},
                 2_{36},4_{41},4_{44},1_{45},4_{56}\}$\cr
\+$8_8^2$& $34$&  $+2$&$[-1,34,34]\{1_0,2_4,2_8,2_{15},2_{16},2_{19},2_{32},
      2_{35},2_{36},2_{43},$\cr
\+&&&$  2_{47},1_{51},2_{52},2_{55},
    2_{59},2_{60},2_{64},2_{67}\}$\cr
\+$8_9^2$& $28$& $-4$&$[1,28,28]\{1_0,4_1,2_4,2_8,4_9,2_{16},4_{25},1_{28},
                            2_{32},2_{36},2_{44},2_{49}\}$\cr
\+&&$0$&$[-1,28,28]\{2_2,4_{11},1_{14},2_{18},2_{22},2_{30},2_{35},1_{42},
                                 4_{43},2_{46},2_{50},4_{51}\}$\cr
\+$8_{10}^2$&  $32$&
$0$&$[-1,8,32]\{2_0,1_3,1_{11},2_{12},1_{19},1_{27},1_{35},
                            1_{43},2_{44},2_{48},1_{51},1_{59}\}$\cr
\+$8_{11}^2$& $28$&  $0$&$[-1,28,28]\{1_0,2_4,2_8,4_{11},2_{16},1_{28},2_{32},
                      2_{35},2_{36},4_{43},2_{44},4_{51}\}$\cr
\+&&$+8$&$[3,28,28]\{1_0,2_4,2_7,2_8,4_{15},2_{16},4_{23},1_{28},2_{32},
                      2_{36},4_{39},2_{44}\}$\cr
\+$8_{12}^2$& $32$&
$+2$&$[1,8,32]\{2_0,1_3,1_{11},2_{12},1_{19},1_{27},1_{35},
                        1_{43},2_{44},2_{48},1_{51},1_{59}\}$\cr
\+$8_{13}^2$& $40$&
$+2$&$[-1,10,40]\{1_0,2_{11},2_{16},2_{19},1_{35},2_{44},2_{51},
                     2_{59},1_{60},2_{64},1_{75},2_{76}\}$\cr
\+$8_{14}^2$& $36$&
$0$&$[1,36,12]\{1_0,4_4,8_7,4_8,1_{12},2_{15},4_{16},4_{20},
                                     8_{23}\}$\cr
\+&&$+8$&$[-3,36,12]\{1_0,2_3,4_4,4_8,8_{11},1_{12},4_{16},8_{19},4_{20}\}$\cr
\+$8_{15}^2$&  $8$&  $0$&$[-1,2,8]\{1_0,1_7,1_{12},1_{15}\}$\cr
\+$8_{16}^2$&  $12$&    $-4$&$[1,12,12]\{1_0,4_1,2_4,2_9,1_{12},2_{16}\}$\cr
\+&&$+4$&$[-1,12,12]\{2_3,1_6,2_{10},1_{18},4_{19},2_{22}\}$\cr
\+$8_1^3$& $20$&
$-8$&$[2,20,10]\{1_0,2_2,4_3,4_7,2_8,1_{10},2_{12},2_{15},2_{18}\}$\cr
\+&&$0$&$[-2,20,10]\{1_0,2_2,2_5,2_8,1_{10},2_{12},4_{13},4_{17},2_{18}\}$\cr
\+&&$+4$&$[4,20,10]\{2_0,2_3,1_5,2_7,4_8,4_{12},2_{13},1_{15},2_{17}\}$\cr
\+$8_2^3$& $28$&
$-8$&$[4,28,14]\{1_0,4_5,2_6,2_{10},2_{12},4_{13},1_{14},4_{17},
                            2_{20},2_{21},2_{24},2_{26}\}$\cr
\+&&$0$&$[0,28,14]\{1_0,4_3,2_6,2_7,2_{10},2_{12},1_{14},4_{19},2_{20},
                                     2_{24},2_{26},4_{27}\}$\cr
\+&&$+2$&$[-1,28,28]\{2_5,2_7,2_{13},2_{17},1_{21},4_{31},2_{33},2_{41},
                        2_{45},4_{47},1_{49},4_{55}\}$\cr
\+$8_3^3$& $28$&
$-4$&$[2,28,7]\{1_0,6_3,6_5,2_6,3_7,2_{10},2_{12},6_{13}\}$\cr
\+&&$+4$&$[-2,28,7]\{3_0,2_3,2_5,6_6,1_7,6_{10},6_{12},2_{13}\}$\cr
\+&&$+6$&$[-3,28,28]\{2_5,2_{13},6_{17},1_{21},6_{33},6_{41},2_{45},3_{49}\}$\cr
\+$8_4^3$& $32$&
$0$&$[0,8,8]\{1_0,2_2,2_3,1_4,2_5,1_8,2_{11},1_{12},2_{13},2_{14}\}$\cr
\+&&$+8$&$[4,8,8]\{1_0,2_3,1_4,2_5,2_6,1_8,2_{10},2_{11},1_{12},2_{13}\}$\cr
\+$8_5^3$& $32$&
$-1$&$[2,8,16]\{1_0,2_1,2_4,1_8,2_9,2_{12},1_{16},2_{17},1_{24},
                                                      2_{25}\}$\cr
\+&&$+2$&$[0,8,16]\{1_0,2_1,2_4,1_8,2_9,1_{16},2_{17},1_{24},
                               2_{25},2_{28}  \}$\cr
\+&&$+4$&$[-1,8,16]\{2_0,1_4,2_5,1_{12},2_{13},1_{20},2_{21},2_{24},1_{28},
                                            2_{29}\}$\cr
\+$8_6^3$& $36$&
$-4$&$[2,36,6]\{5_0,2_2,10_3,2_4,5_6,4_7,2_8,2_{10},4_{11}\}$\cr
\+&&$0$&$[0,36,6]\{10_0,2_1,5_3,4_4,2_5,2_7,4_8,5_9,2_{11}\}$\cr
\+&&$+4$&$[-2,36,18]\{5_0,4_3,2_6,2_{12},4_{15},5_{18},2_{24},10_{27},2_{30}\}$\cr
\+$8_7^3$&    $4$&   $-4$&$[0,4,2]\{2_0,1_1,1_3\}$\cr
\+&&$0$&$[-2,4,2]\{1_0,1_2,2_3\}$\cr
\+&&$+8$&$[2,4,2]\{1_0,2_1,1_2\}$\cr
\+$8_8^3$&  $12$&    $-4$&$[2,12,6]\{2_0,2_1,1_3,4_4,2_7,1_9\}$\cr
\+&&$0$&$[0,12,6]\{1_0,4_1,2_4,1_6,2_9,2_{10}\}$\cr
\+&&$+8$&$[4,12,6]\{1_0,2_3,2_4,1_6,4_7,2_{10}\}$\cr
\+$8_9^3$&   $16$&   $-4$&$[2,4,4]\{1_0,2_1,2_2,1_4,2_5\}$\cr
\+&&$0$&$[0,4,4]\{2_0,1_2,2_3,1_6,2_7\}$\cr
\+&&$+4$&$[-2,4,4]\{1_0,2_1,1_4,2_5,2_6\}$\cr
\+$8_{10}^3$&  $0$&  $-4$&$[3,4,4]\{1_2,2_3,1_6\}, \hskip 0.5cm \ell =1$\cr
\+&&$0$&$[1 ,4,4]\{1_{0},2_{1},1_{4}\}, \hskip 0.5cm \ell =1$\cr
\+&&$+4$&$[-3,4,4]\{1_0,1_4,2_5\}, \hskip 0.5cm \ell =1$\cr
\+$8_1^4$&  $32$&    $-8$&$[3,8,8]\{1_0,4_3,3_4,3_8,4_{11},1_{12}\}$\cr
\+&&$-4$&$[1,8,8]\{3_0,3_4,4_7,1_8,1_{12},4_{15}\}$\cr
\+&&$0$&$[-1,8,8]\{3_0,4_3,1_4,1_8,4_{11},3_{12}\}$\cr
\+$8_2^4$& $16$&    $-4$&$[1,4,4]\{2_0,3_1,2_4,1_5\}$\cr
\+&&$-4$&$[-1,4,4]\{2_2,1_3,2_6,3_7\}$\cr
\+&&$+4$&$[-3,4,4]\{2_0,1_1,2_4,3_5\}$\cr
\+$8_3^4$& $0$&    $0$&$[0,4,1]\{3_0,1_1\}, \hskip 0.5cm \ell =1$\cr
\+&&$+8$&$[4,4,1]\{1_0,3_1\}, \hskip 0.5cm \ell =1$\cr

\endpage

\centerline {\bf References}

\noindent
[1]  V. F. R. Jones,
 {\it Pacific J. Math.} {\bf 137}:311 (1989).
\nextline
[2] M.  Wadati, T.  Deguchi, and Y. Akutsu,
 {\it Phys. Rep.} {\bf 180}:247 (1989).
\nextline
[3] F. Y. Wu, {\it Rev. Mod. Phys.}
{\bf 64}:1099 (1992).
\nextline
[4] V. F. R. Jones,
 {\it Bull. Amer. Math. Soc.} {\bf 12}:103 (1985).
\nextline
[5] P.  Freyd, D. Yetter, J. Hoste, W. B. R. Lickorish, K. C. Millett,
and A. Oceanau, {\it Bull. Am. Math. Soc.} {\bf 12}:239 (1985).
\nextline
[6] D. Goldschmidt and V. F. R. Jones, {\it Geom. Dedicata}
{\bf 31}:165 (1989);  V. F. R. Jones, {\it Commun. Math. Phys.}
{\bf 125}:459 (1989).
\nextline
[7] E. Date, M. Jimbo, K. Miki and T. Miwa,
Pacific J. Math. {\bf 154}:37 (1992).
\nextline
[8] F. Jaeger, {\it Geom. Dedicata} {\bf 44}:23 (1992).
\nextline
[9] P. de la Harpe and V. F. R. Jones,
{\it J. Comb. Theory} {\bf B57}:207 (1993).
\nextline
[10] E. Bannai and E. Bannai, in {\it Memoirs of the Faculty of
Science}, Kyushu University, {\bf A47}:397 (1993).
\nextline
[11] P. de la Harpe, {\it Pacific J. Math.} {\bf 162}:57 (1994).
\nextline
[12] F. Y. Wu, P. Pant, and C. King, {\it Phys. Rev. Lett.} {\bf 72}:3937
(1994).
\nextline
[13] H. Au-Yang, B. M. McCoy, J. H. H. Perk, S. Tang, and M. L. Yan,
{\it Phys. Lett.} {\bf 123A}:219 (1987).
\nextline
[14] R. J. Baxter, J. H. H. Perk, and H. Au-Yang, {\it Phys. Lett.}
{\bf A128}:138 (1988).
\nextline
[15] C. L. Siegel, {\it Nachr. der Akad. Wiss. G\"ottingen Math.-Phys.
Klasse}
{\bf 1}:1 (1960).
\nextline
[16] T. Kobayashi, H. Murakami and J. Murakami, {\it Proc.
Japan Acad.} {\bf 64A}:235 (1988).
\nextline
[17] R. J. Baxter, S. B. Kelland, and F. Y. Wu, {\it J. Phys.}
A{\bf 9}:397 (1976).
\nextline
[18] K. Reidemeister,  {\it Knotentheorie} (Chelsea, New York, 1948).
\nextline
[19] F. Y. Wu and Y. K. Wang, {\it J. Math. Phys.} {\bf 17}:439 (1976).
\nextline
[20] V. A. Fateev and A. B. Zamolodchikov, {\it
Phys. Lett.} {\bf A92}:37 (1982).
\nextline
[21] F. Harary,  {\it Graph Theory}, (Addison-Wesley, New York, 1971).
\nextline
[22] L. Goeritz, {\it Math. Z.} {\bf 36}:647 (1933).
\nextline
[23] G. Burde and H. Zieschang, {\it Knots}, (Walter de Gruyter, New York,
1985).
\nextline
[24]  L. A.  Traldi,{\it Math. Z.} {\bf 188}:203 (1985).
\nextline
[25]  Y. Akutsu and M. Wadati, {\it J. Phys. Soc. Japan} {\bf 56}, 3039 (1987).
\nextline
[26] W. B. R. Lickorish and K. C. Millett, {\it Math. Magazine} {\bf 61}:3
(1988).
\nextline
[27] See, for example, J. W. S. Cassels, {\it Rational Quadratic Forms},
(Academic Press, London, 1972).
 \nextline
[28] C. F. Gauss, {\it Summatio quarundam serierum singularium, 1808},
       in {\it Werke} II (G\"ottingen, 1870).
\nextline
[29] A. Krazer, in {\it Festschr. H. Weber} (Leipzig u., Berlin 1912).
\nextline
[30] M. Eichler, {\it Quadratische Formen und Orthogonale Gruppen},
(Springer-
\nextline
Verlag, Berlin, 1952).
\nextline

\endpage

\centerline {Figure captions}

\noindent
FIG. 1. The $+$ and $-$  line intersections associated with
two kinds of
shadings that can occur at a vertex and the vertex weights.

\noindent
FIG. 2. The two different kinds of face shadings
 and the associated line graphs for the link
$8^2_{15}$, where the interaction types are explicitly noted.

\noindent
FIG. 3. Reidemeister moves for oriented knots with two different
kinds of face shadings.

\noindent
FIG. 4. Auxiliary lines defining rapidities for the chiral Potts model
and the vertex weights.

\noindent
FIG. 5. Example of a shaded polygon showing one
source and one sink of arrows around its perimeter.

\noindent
FIG. 6. Signed graphs correspond to the
two possible face shadings of the link $8^2_{15}$ shown in Fig. 2.

\noindent
FIG. 7. Line configurations that can occur at a line
crossing for writing down the  Skein relation.

\noindent
FIG. 8. Consideration of split and connected links.

\noindent
FIG. 9. (a) A tangle cut from a link $K$. (b) The tangle
with two half twists added to its lines to be reconnected
to $K$.

\bye